\documentstyle[12pt]{article}
\newlength{\pubnumber} 

\catcode`\@=11
\@addtoreset{equation}{section}

\def\section{\@startsection{section}{1}{\z@}{3.5ex plus 1ex minus .2ex}
 {2.3ex plus .2ex}{\large\bf}}
\def\subsection{\@startsection{subsection}{2}{\z@}{2.3ex plus .2ex}
 {2.3ex plus .2ex}{\bf}}

\begin{document}

\begin{titlepage}
\samepage{
\setcounter{page}{1}
\vskip 5.5truecm
\begin{center}
  {\large\bf Character States and Generator Matrix
   Elements for $Sp(4) \supset SU(2) \times U(1)$\\}

 \vskip 1.5truecm
     {\bf N. Hambli, J. Michelson and R. T. Sharp \\}
   {Department of Physics, McGill University\\
    3600 University St., Montr\'eal, Qu\'ebec~H3A-2T8~~Canada}
\footnote{Research supported in part by the Natural Sciences
and Engineering Research Council of Canada and by the Fonds
FCAR du Qu\'ebec.}\\
\vskip 2truecm
\today\\
\end{center}
\vfill\eject
\begin{abstract}
   {A new set of polynomial states (to be called character
    states) are derived for $Sp\,(4)$ reduced to its
    $SU(2) \times U(1)$ subgroup, and the relevant generator
    matrix elements are evaluated. The group--subgroup in
    question is that of the seniority model of nuclear physics.\\}
\end{abstract}
 }
\end{titlepage}

\setcounter{footnote}{0}
\def\beq{\begin{equation}}
\def\eeq{\end{equation}}
\def\beqn{\begin{eqnarray}}
\def\barrl{\begin{array}{ll}}
\def\eeqn{\end{eqnarray}}
\def\earr{\end{array}}


\setcounter{footnote}{0}
\section{ Introduction}

The group--subgroup $Sp(4) \supset SU(2) \times U(1)$
finds application in the nuclear seniority model
(Flowers 1952). Polynomial basis states (in the states
of the $Sp(4)$ fundamental representations) have been
given by Hecht (1965), Parikh (1965), Ahmed and Sharp
(1970) and by Smirnov and Tolstoy (1973). In this paper
we present new polynomial basis states, which we christen
``character states''; we believe then to be simpler than
any of the earlier ones. Analogous states have been given
recently for $SU(3)$ and $SO(5)$ reduced according to
their finite Demazure--Tits subgroup (de Guise and
Sharp 1991), for $SO(7)$ reduced according to $SU(2)^3$
(Burdik, Cummins, Gaskell and Sharp 1992) and for
$G_2$ reduced according to $SU(3)$ (Farell, Lam
and Sharp 1993).

In Section~(2) we discuss character states in the
context of $Sp(4) \supset SU(2) \times U(1)$.
Section~(3) deals with generator matrix elements
for the degenerate representations $(a,0)$ and
$(0,b)$. Section~(5) contains some concluding
remarks.

\setcounter{footnote}{0}
\section{Character States For $Sp(4) \supset SU(2) \times U(1)$}

The basis states of the $Sp(4)$ representation
$(a,b)$ are polynomials of degrees $a,b$ in the
states of the respective fundamental representations.
Hence only stretched IR's (representation labels
additive) in the direct product of $a$ copies of $(1,0)$
and $b$ copies of $(0,1)$ are to be retained.

Consider the quadratic direct products
\begin{eqnarray}
{(1,0)^2}_{10} & = & (2,0)_{10}\; ,\nonumber\\
{(0,1)^2}_{15} & = & (0,2)_{14} + (0,0)_{1}\; ,
\label{aa}\\
(1,0) \times (0,1)_{20} & = & (1,1)_{16} + (1,0)_{4}\; .
\nonumber
\end{eqnarray}
The square of a representation above means the symmetric
(polynomial) part of the direct product of two copies.
A subscript on a representation or product is its dimension.
The stretched, or wanted, part of each product is the first
representation on the right; the states of the other
representations are unwanted for the purpose of forming
our polynomial basis.

The $Sp(4)$ character generator may be written
\begin{equation}
X(A,B;M,Z) =
{1\over{\beta \eta \zeta}}\;
\left[{1\over{\xi \alpha \theta}} +
{\gamma\over{\alpha \theta \gamma}} +
{\kappa\over{\theta \gamma \kappa}} +
{\delta\over{\gamma \kappa \delta}} +
{{\alpha\delta}\over{\gamma \delta \alpha}}\right]\; ;
\label{ab}
\end{equation}
we have adopted the space--saving convention that a
variable in a denominator stands for unity minus that
variable. To stress our interpretation of the character
generator as generating function for basis states we have
expressed $X$ in terms of the fundamental representation
states (see figure 1). In the power series expansion of
(\ref{ab}) each term of degree $a$ in the $(1,0)$
variables and degree $b$ in the $(0,1)$ variables
represents one state of the representation $(a,b)$.
The states thus defined are complete and non
redundant. We call them character states. We remark
that they are contaminated by unwanted states belonging
to lower representations; that does not matter for the
purpose of computing generator matrix elements. For $X$
as a generating function for characters, or weights,
the following substitutions should be made:
\begin{eqnarray}
\alpha & \to & A M^{1\over2} Z^{1\over2},\quad
\beta  \to  A M^{1\over2} Z^{-\, {1\over2}},\quad
\gamma  \to  A M^{-{1\over2}} Z^{1\over2}, \quad
\delta  \to  A M^{-{1\over2}} Z^{-{1\over2}}\; ,
\nonumber\\
\eta & \to & B Z,\quad
\xi  \to  B M~,\quad
\zeta  \to  B Z^{-1}, \quad
\kappa  \to  B M^{-1}~, \quad\
\theta  \to  B\; .
\label{ac}
\end{eqnarray}
The variables $A$ and $B$ are dummies which carry the
$Sp(4)$ representation labels $a$ and $b$ as exponents;
$M$ carries the $SU(2)$ weight $m$ and $Z$ the $U(1)$
label $z$. Then in the expansion of $X$,
\begin{eqnarray}
X(A,B;M,Z) = \sum_{abmz}\; A^a B^b M^m Z^z C_{abmz}\; ,
\label{ad}
\end{eqnarray}
the coefficient $C_{abmz}$ is the multiplicity of the
weight $(m,z)$ in the IR $(a,b)$. The character generator
(\ref{ab}) with the substitutions (\ref{ac}) agrees with
earlier versions (Gaskell, Peccia and Sharp 1978, Patera
and Sharp 1979, de Guise and sharp 1991) when the weights
are expressed in the same basis and the terms put over a
common denominator.

Examination of (\ref{ab}) reveals that certain pairs
of variables, namely $\alpha\kappa$, $\gamma\xi$,
$\delta\xi$, $\delta\theta$, $\xi\kappa$ never
appear in the same term and therefore these
products never appear in the expression for a
state; we say they are incompatible.
Each incompatible pair appears as one term in
the expression for one of the unwanted states
on the right hand side of (\ref{aa}).
Setting the unwanted states equal to zero
and solving for the incompatible pairs gives
the following substitutions by which incompatible
pairs may be eliminated in favour of compatible ones
when they arise in the course of a calculation:
\begin{eqnarray}
\alpha\, \kappa & = &  \delta \eta +
{\sqrt{2}\over 2}\; \gamma \theta\; ,
\nonumber\\
\gamma\, \xi  & = & -\; \beta \eta +
{\sqrt{2}\over 2}\; \alpha \theta\; ,
\nonumber\\
\delta\, \xi  & = &  \alpha \zeta +
{\sqrt{2}\over 2}\; \beta \theta\; ,
\label{ae}\\
\delta\, \theta & = &  \sqrt{2}\; \beta \kappa +
\sqrt{2}\; \gamma \zeta\; ,
\nonumber\\
\xi\, \kappa & = & \eta \zeta + {1\over 2}\; \theta^2\; .
\nonumber
\end{eqnarray}

It is shown by Farell, Lam and Sharp 1994, and Hambli and Sharp
1995 that elementary unwanted states are all of degree $2$ and
hence that elementary incompatibilities are between pairs of
states only, a fact verified straightforwardly for $Sp(4)$.

We complete this section by giving the
$Sp(4) \supset SU(2) \times U(1)$ branching
rules generating function (Sharp and Lam 1969 give
the equivalent integrity basis):
\begin{equation}
{1\over{\left(1 - A T^{1\over 2} Z^{1\over 2}\right)
\left(1 - A T^{1\over 2} Z^{- {1\over 2}}\right)
\left(1 - B Z\right)
\left(1 - B Z^{-1}\right)}}
\left[
{1\over{\left(1 -  B T\right)}} +
{{A^2}\over{\left(1 - A^2\right)}}
\right]\; .
\label{af}
\end{equation}
The elements of (\ref{af}) may be interpreted as
the highest states of subgroups IR's contained in
low $Sp(4)$ IR's:
\begin{eqnarray}
\alpha & \sim & A T^{1\over2} Z^{1\over2},\quad
\beta  \sim  A T^{1\over2} Z^{-\, {1\over2}},\quad
\alpha\delta - \beta\gamma  \sim  A^2 \; ,
\nonumber\\
\eta & \sim & B Z,\quad
\zeta  \sim  B Z^{- 1},\quad
\xi  \sim  B T \; .
\label{ag}
\end{eqnarray}
Thus the highest state of a subgroup IR
may be written, in unnormalized form, and
representation labels suppressed,
\begin{equation}
\left|
\, t\;,t\;,z\; ; v\,
\right\rangle =
\alpha^x \; \beta^y\; \eta^u\;
\zeta^v\; \xi^w\;
\left(\alpha\delta - \beta\gamma\right)^s\; ,
\label{ah}
\end{equation}
with
\beqn
a & = & x + y + 2s ~ , \quad
b  =  u + v + w \; ,
\nonumber\\
& &\label{ai}\\
t & = & {1\over 2}\; \left(x + y\right) + w~ , \quad
z   =  {1\over 2}\; \left(x - y\right) + u - v \; ,
\nonumber
\eeqn
and $s\; w = 0$. We allow $v$ to play the role of the
missing label.

We may distinguish two types of state, according
to whether (type I) $s = 0$ (and $t \ge a/2$)
or (type II) $w = 0$ (and $t \le a/2$). Solving
(\ref{ai}) for $x$, $y$, $u$ and $w$ or $s$,
(\ref{ah}) becomes
\begin{equation}
\left|
\,t\;,t\;,z\; ; v\,
\right\rangle =
\alpha^{t + z - b + 2 v} \; \beta^{a + b - t - z - 2 v} \;
\eta^{{a\over 2} + b - t - v} \;
\zeta^v\;
\xi^{t - {a\over 2}} \; ,
\label{aj}
\end{equation}
for type I, and, for type II
\begin{equation}
\left|
\,t\;,t\;,z\; ; v\,
\right\rangle =
\alpha^{t + z - b + 2 v} \;
\beta^{b + t - z - 2 v}\;
\left(\alpha\delta - \beta\gamma\right)^{{a\over 2} - t}\;
\eta^{b - v}\;
\zeta^v\; .
\label{ak}
\end{equation}
For $t = a/2$ the two types of state coincide.

The $SU(2)$ root generators are
\begin{eqnarray}
T_{+} & = &  \alpha\, \partial_{\gamma} +
\beta \partial_{\delta} +
\sqrt{2}\, \left(\xi\, \partial_\theta +
\theta\, \partial_\kappa \right)\; ,
\nonumber\\
& & \label{al}\\
T_{-}  & = &  \gamma\, \partial_{\alpha} +
\delta \partial_{\beta} +
\sqrt{2}\, \left(\theta\, \partial_\xi +
\kappa\, \partial_\theta \right)\; .
\nonumber
\end{eqnarray}
The other $6$ $Sp(4)$ root generators
constitute an $SU(2)$ vector $G$ with
$z = 1$,
\begin{eqnarray}
G_{+1} & = &  \alpha\, \partial_\delta +
\eta\, \partial_\kappa + \xi\, \partial_\zeta \; ,
\nonumber\\
G_{-1}  & = &  -\; \gamma\, \partial_\beta +
\kappa \partial_\zeta +
\eta\, \partial_\xi \; ,
\label{am}\\
G_{0} & = & {\sqrt{2}\over 2}\,
\left(\gamma\, \partial_\delta -
\alpha\, \partial_\beta \right) +
\theta\, \partial_\zeta -
\eta\, \partial_\theta \; ,
\nonumber
\end{eqnarray}
and a vector $\overline{G}$ with $z = - 1$,
\begin{eqnarray}
\overline{G_{+1}} & = &  \beta\, \partial_\gamma -
\xi\, \partial_\eta - \zeta\, \partial_\kappa \; ,
\nonumber\\
\overline{G_{-1}}  & = &  -\; \delta\, \partial_\alpha -
\kappa \partial_\eta - \zeta\, \partial_\xi \; ,
\label{an}\\
\overline{G_{0}} & = & {\sqrt{2}\over 2}\,
\left(\delta\, \partial_\gamma  -
\beta\, \partial_\alpha \right) +
\zeta\, \partial_\theta -
\theta\, \partial_\eta \; .
\nonumber
\end{eqnarray}
$G$ and $\overline{G}$ are hermitian conjugate  with
\begin{eqnarray}
\overline{G_{i}} = \left(- 1\right)^i\;\;
G_{- i}^{\dagger}\; .
\label{ao}
\end{eqnarray}

We will save considerable work later by noticing
that under the substitutions
$\alpha \leftrightarrow \delta$,
$\beta \leftrightarrow \gamma$,
$\eta \leftrightarrow \zeta$,
$\xi \leftrightarrow \kappa$,
we have
$G_i \leftrightarrow -\, \overline{G_{-i}}$
and that under the same substitutions we have
for the type I states (\ref{aj})
$\left|\,t\;, m\;, z\; ; v\,\right\rangle
\leftrightarrow
\left|\,t\;, -m\;, -z\; ; a/2 + b - t - v\,\right\rangle$
while for type II states (\ref{ak})
$\left|\,t\;, m\;, z\; ; v\,\right\rangle
\leftrightarrow
\left|\,t\;, -m\;, -z\; ; b - v\,\right\rangle$.

\setcounter{footnote}{0}
\section{Generator Matrix Elements For Degenerate
Representations}

For the degenerate IR's $(a,0)$ and $(0,b)$ there
is no missing label and the basis states are orthogonal
and hence can be normalized straightforwardly. Since
in that sense they can be handled more satisfactorily
than generic states $(a > 0 , b > 0)$ we treat them
separately.

For $(a,0)$, according to (\ref{ak}) with $v = 0$, the
highest state of an $SU(2)$ IR is (we suppress the
label $a$)
\begin{equation}
\left|
\,t\;,t\;,z\; \,
\right\rangle = N_{t\;\;z}\;\;
\alpha^{t + z}\; \beta^{t - z}\;
\left(\alpha\delta - \beta\gamma\right)^{{a\over 2} - t}\; .
\label{ba}
\end{equation}
The branching rule is $a/2 \ge t \ge |z|$ with
$2 t$ and $2 z$ having the parity of $a$.

$(a,0)$ states are not contaminated by unwanted states
and can be normalized by standard methods.
For brevity write
\begin{equation}
\left|\, h \,\right\rangle = N_{h}\;\;
\alpha^{f}\; \beta^{g}\;
\left(\alpha\delta - \beta\gamma\right)^{h}\; ,
\label{bb}
\end{equation}
and equate
\begin{equation}
\left\langle\, h + 1\,\right|\,
\alpha\, \delta - \beta \,\gamma \,
\left|\, h \,\right\rangle =
\left\langle\, h \,\right|\,
\partial_\alpha \, \partial_\delta -
\partial_\beta \, \partial_\gamma \,
\left|\, h + 1\,\right\rangle \; ,
\label{bc}
\end{equation}
to obtain a recursion relation for $N_h$,
whose solution is, for $N_{t\;\;z}$,
\begin{equation}
N_{t\;\;z} =
\sqrt{
{{(2 t + 1)!}
\over
{\left(t + z\right) !\; \left(t - z\right) !\;
\left({a\over 2} - t\right) !\;
\left({a\over 2} + t + 1 \right) !}}
}\; .
\label{bd}
\end{equation}

We can write
\begin{equation}
G_0 \; \left|\, t\; , t\; , z\,\right\rangle = \;
A\; \left|\, t+1\; , t\; , z+1\,\right\rangle +
B\; \left|\, t\; , t\; , z+1\,\right\rangle \; ,
\label{bd}
\end{equation}
where the matrix elements $A$ and $B$ will now
be determined. Apply $T_{+}$ to
(\ref{bd}) with the result
\begin{equation}
\sqrt{2}\; G_1 \; \left|\, t\; , t\; , z\,\right\rangle = \;
\sqrt{2(t+1)}\; A\; \left|\, t+1\; , t+1\; , z+1\,\right\rangle \; ,
\label{be}
\end{equation}
from which
\begin{eqnarray}
A & = &
\left\langle\, t+1\;,t\;,z+1\,\right|\,
G_0 \,
\left|\, t\;,t\;,z \,\right\rangle
= {{{a\over 2} - t}\over{\sqrt{t+1}}}\;\;
{{N_{t\;\;z}}\over{N_{t+1\;\;z+1}}}\; ,
\nonumber\\
& = &
{1\over{t + 1}}\;\;
\sqrt{
{{\left(t + z + 1\right) \left(t + z + 2\right)
\left({a \over 2} - t\right) \left({a \over 2} + t + 2\right)}
\over
{2\, \left(2 t + 3\right)}}
}\; .
\label{bf}
\end{eqnarray}

Inserting this value for $A$ in (\ref{bd}) we obtain
for $B$
\begin{eqnarray}
B & = &
\left\langle\, t\;,t\;,z+1\,\right|\,
G_0 \,
\left|\, t\;,t\;,z \,\right\rangle
= -\; {{({a\over 2} + 1)(t - z)}\over{\sqrt{2} (t + 1)}}\;\;
{{N_{t\;\;z}}\over{N_{t\;\;z+1}}}\; ,
\nonumber\\
& = &
-\; {{{a\over 2} + 1}\over{\sqrt{2} (t + 1)}}\;\;
\sqrt{
\left(t + z + 1\right) \left(t - z\right)
}\; .
\label{bg}
\end{eqnarray}

To obtain a matrix element of $G_0$ in which $G_0$
reduces $t$ by unity make the following transformations:
\begin{eqnarray}
\left\langle t+1\;,t\;,z+1\right|
G_0
\left| t\;,t\;,z \right\rangle
& = & -
\left\langle t+1\;,-t\;,-z-1\right|\,
\overline{G_0}
\left|\, t\;,-t\;,-z \,\right\rangle\; ,
\nonumber\\
& = &
-
\left\langle t\;,-t\;,-z\right|
G_0
\left| t+1\;,-t\;,-z-1 \right\rangle \, .
\label{bh}
\end{eqnarray}
The first step follows from the substitutions in
the paragraph following (\ref{ao}) and the second
is a consequence of (\ref{ao}).

For the reduced matrix element of $G$ we get
\begin{eqnarray}
\left\langle \,t+1\;,z+1\,
\left|\!\left| \,G\, \right|\!\right|
\,t\;,z\, \right\rangle & = &
\sqrt{
{{\left(t + z + 1\right) \left(t + z + 2\right)
\left({a\over 2} - t\right) \left({a\over 2} + t + 2\right)}
\over
{2 \left(t + 1\right)}}
}\; ,
\nonumber\\
\left\langle \,t\;,z+1\,
\left|\!\left| \,G\, \right|\!\right|
\,t\;,z\, \right\rangle & = &
-\; \left({a\over 2} + 1\right)\;
\sqrt{
{{\left(2 t + 1\right) \left(t + z + 1\right) \left(t - z\right)}
\over
{2 t \left(t + 1\right)}}
}\;\; ,
\label{bi}\\
\left\langle \,t-1\;,z+1\,
\left|\!\left| \,G\, \right|\!\right|
\,t\;,z\, \right\rangle & = &
\sqrt{
{{\left(t - z - 1\right) \left(t - z\right)
\left({a\over 2} - t + 1\right) \left({a\over 2} + t + 1\right)}
\over
{2 t}}
}\;\; .
\nonumber
\end{eqnarray}
Because of (\ref{ao}) the reduced matrix elements of
$\overline{G}$ are given in terms of those of $G$
by the relation
\begin{equation}
\left\langle \,t+k\;,z-1\,
\left|\!\left| \,\overline{G}\,
\right|\!\right|\,t\;,z\, \right\rangle =
-\left(-1\right)^k \,
\left\langle \,t\;,z\,
\left|\!\left| \,G\, \right|\!\right|
\,t+k\;,z-1\, \right\rangle\; .
\label{bip}
\end{equation}

We turn to $(0,b)$ states. According to (\ref{aj})
with $v =(b - t - z)/2$ the highest state of a subgroup
multiplet is
\begin{equation}
\left| \, t\; , t \; , z \,\right\rangle =
N_{t\;\;z} \,
\eta^{{(b-t+z)}\over 2} \,
\zeta^{{(b-t-z)}\over 2} \,
\xi^t\;\; .
\label{bj}
\end{equation}
The branching rule is $b - |z| \ge t \ge 0$.
The $SU(2)$ ladder generators are now
\begin{eqnarray}
T_{+} & = &
\sqrt{2}\; \left(
\xi\, \partial_\theta + \theta\, \partial_\kappa
\right)\; ,
\nonumber\\
& &\label{bk}\\
T_{-} & = &
\sqrt{2}\; \left(
\theta\, \partial_\xi + \kappa\, \partial_\theta
\right)\; .
\nonumber
\end{eqnarray}
The components of $G$ and $\overline{G}$ are
\begin{eqnarray}
G_{+1} & = &  \eta\, \partial_\kappa +
\xi\, \partial_\zeta \; ,
\nonumber\\
G_{-1}  & = &   \kappa\, \partial_\zeta +
\eta \partial_\xi \; ,
\label{bl}\\
G_{0} & = &
\theta\, \partial_\zeta -
\eta\, \partial_\theta \; ,
\nonumber
\end{eqnarray}
\begin{eqnarray}
\overline{G_{+1}} & = &  -\, \xi\, \partial_\eta -
\zeta\, \partial_\kappa \; ,
\nonumber\\
\overline{G_{-1}}  & = &  -\; \kappa\, \partial_\eta -
\zeta \partial_\xi \; ,
\label{bm}\\
\overline{G_{0}} & = &
\zeta\, \partial_\theta -
\theta\, \partial_\eta \; .
\nonumber
\end{eqnarray}
Again $\overline{G_i} = \left(-1\right)^i \;
{G_{-i}}^\dagger$. The basis states are contaminated
by states containing as a factor the unwanted scalar
$\eta\;\zeta - \xi\;\kappa + \theta^2 /2$.

Applying  $\overline{G_{-1}}$ to
$\left|\, t\; ,t\; , z\, \right\rangle$ yields
\begin{equation}
\overline{G_{-1}} \; \left|\, t\; , t\; , z\,\right\rangle = \;
A\; \left|\, t+1\; , t-1\; , z-1\,\right\rangle +
C\; \left|\, t-1\; , t-1\; , z-1\,\right\rangle \; .
\label{bn}
\end{equation}
($t$ changes only by $\pm 1$ because it has the
parity of $b+z$.) We must now find the matrix elements
$A$ and $C$. Application of $T_{+}^2$ to
(\ref{bn}) gives
\begin{equation}
\overline{G_{1}} \; \left|\, t\; , t\; , z\,\right\rangle = \;
A\; \sqrt{(t+1)(2t+1)} \;
\left|\, t+1\; , t+1\; , z-1\,\right\rangle \; ,
\label{bo}
\end{equation}
from which
\begin{eqnarray}
A & = &
\left\langle\, t+1\;,t-1\;,z-1\,\right|\,
\overline{G_{-1}} \,
\left|\, t\;,t\;,z \,\right\rangle \; ,
\nonumber\\
& = & -\, {1\over 2}\;
{{\left(b-t+z\right)}\over{\sqrt{(t+1)(2t+1)}}}\;\;
{{N_{t\;\;z}}\over{N_{t+1\;\;z-1}}}\; .
\label{bp}
\end{eqnarray}
Then (\ref{bn}) yields immediately
\begin{eqnarray}
C & = &
\left\langle\, t-1\;,t-1\;,z-1\,\right|\,
\overline{G_{-1}} \,
\left|\, t\;,t\;,z \,\right\rangle \; ,
\nonumber\\
 & = & -\,
{{t(b+t+z+1)}\over{2t+1}}\;\;
{{N_{t\;\;z}}\over{N_{t-1\;\;z-1}}}\; .
\label{bq}
\end{eqnarray}
Applying $G_{1}$ to $\left|\,t-1\;,t-1\;,z-1\,\right\rangle$
gives
\begin{equation}
\left\langle\, t\;,t\;,z\,\right|\,
{G_{1}} \,
\left|\, t-1\;,t-1\;,z-1 \,\right\rangle
= {1\over 2}\;(b - t - z + 2)\;\;
{{N_{t-1\;\;z-1}}\over{N_{t\;\;z}}}\;\; .
\label{br}
\end{equation}
Because $\overline{G_{-1}} = -\; {G_1}^\dagger$ we can
equate the right hand side of (\ref{br}) to the negative
of the right hand side of (\ref{bq}), getting
a recursion relation for the normalization constant
\begin{equation}
{{N_{t\;\;z}}\over{N_{t-1\;\; z-1}}}  =
\sqrt{
{{(b-t-z+2)(2t+1)}\over{2(b+t+z+1)t}}
}\; \; ,
\label{bs}
\end{equation}
whose solution is
\begin{equation}
N_{t\;\;z} =
\phi \;\;
\sqrt{
{(2t+1)!!}
\over
{2^{{1\over 2}\, (b+t-z)} \;
t!\;\;
\left({{b-t-z}\over2}\right)!\;\;
(b+t+z+1)!!}
}\;\; ,
\label{bt}
\end{equation}
where $\phi$ is constant when $t$ and $z$
are given the same increment. Increasing $t$
and $z$ each by $(b-t-z)/2$ takes us to the state
\begin{equation}
\left|
{1\over 2} (b+t-z)\;,
{1\over 2} (b+t-z)\;,
{1\over 2} (b-t+z)
\right\rangle =
N_{{1\over 2} (b+t-z)\;\; {1\over 2}\, (b-t+z)}\;
\eta^{{1\over 2} (b-t+z)}\;
\xi^{{1\over 2} (b+t-z)}\; ,
\label{bu}
\end{equation}
which is on the boundary of the $(0,b)$ weight
diagrams and uncontaminated. It is easily normalized:
\begin{equation}
N_{{1\over 2}\, (b+t-z)\;\; {1\over 2}\, (b-t+z)} =
{1\over{\sqrt{
\left({{b-t+z}\over 2}\right)! \;\;
\left({{b+t-z}\over 2}\right)!
}}}\;\; .
\label{bv}
\end{equation}
Using (\ref{bv}) in (\ref{bt}) determines
$\phi$ to be
\begin{equation}
\phi =
\sqrt{
{{2^{{1\over 2}\, (b+t-z)} \;\; (2b+1)!!}
\over
{\left({{b-t+z}\over 2}\right)! \;\;
\left(b+t-z+1\right)!!}}
}\; \;\; .
\label{bw}
\end{equation}
(\ref{bt}) then gives for $N_{t\;\;z}$
\begin{equation}
N_{t\;\;z} =
\sqrt{
{{(2t+1)!!\;\; (2b+1)!!}
\over
{t!\;\;
\left({{b-t-z}\over 2}\right)!\;\;
\left({{b-t+z}\over 2}\right)!\;\;
(b+t+z+1)!!\;\; (b+t-z+1)!!}}
}\;\;\; .
\label{bx}
\end{equation}
We have normalized the wanted part of the state
(\ref{bj}) without ever isolating it.
As a check consider (\ref{bj}) with
$b=4,t=0,z=0$, {\it i.e.,}
$N_{00}\; \eta^2 \; \zeta^2$.
The wanted part of $\eta^2\;\zeta^2$,
{\it i.e.,} the part orthogonal to
$\eta\zeta
\left(\eta \zeta - \xi \kappa +
{1/ 2}\, \theta^2 \right)$ and to
$\left(\eta \zeta - \xi \kappa +
{1/ 2}\, \theta^2 \right)^2$ is
$1/63\; \left(
15\, \eta^2 \, \zeta^2 +
8\, \xi^2 \, \kappa^2 +
2\, \theta^4 +
40\, \eta\, \zeta\, \xi\, \kappa -
20\, \eta\, \zeta\, \theta^2 -
8\, \xi\, \kappa\, \theta^2\right)$
whose norm is $20/21$ which checks with
(\ref{bx}) by which $N_{00} = \sqrt{21/20}$.

Inserting the value of the normalization constant
in (\ref{bp}) and (\ref{bq}) gives the matrix elements
explicitly. For the reduced matrix elements of $G$
we find
\begin{eqnarray}
\left\langle \,t+1\;,z-1\,
\left|\!\left| \,
\overline{G}\, \right|\!\right|
\,t\;,z\, \right\rangle & = &
-\,\sqrt{
(t+1)\,
\left({{b-t+z}\over 2}\right)\,
(b+t-z+3)}\; ,
\nonumber\\
&  &
\label{by}\\
\left\langle \,t-1\;,z-1\,
\left|\!\left| \,
\overline{G}\, \right|\!\right|
\,t\;,z\, \right\rangle & = &
-\,\sqrt{
t\,
\left({{b-t-z+2}\over 2}\right)\,
(b+t+z+1)}\;\; .
\nonumber
\end{eqnarray}

\section{ Concluding Remarks}

We have defined the matrix elements of a
generator $G$ between states
$\left|\, h\, \right\rangle$ and
$\left|\, g\, \right\rangle$ of a
complete non--redundant set as
$\left(\, g\, \right|\, G\, \left|\, h\, \right)=$
coefficient of $\left|\, g\, \right\rangle$
in $G\, \left|\, h\, \right\rangle$. As long as
the states are orthonormal this is equivalent
to the usual definition
$\left\langle\, g\, \right|\, G\,
\left|\, h\, \right\rangle=$ overlap of
$\left|\, g\, \right\rangle$ with
$G\, \left|\, h\, \right\rangle$; this is the case
when the subgroup used to define the states provides
a complete set of labels, as for the states we deal with
in this paper.

The basis states are defined as the `wanted' parts
of products of powers of mutually compatible sets
of basis states of the fundamental representations.

In a forthcoming paper we will treat generic
representations $(a,b)$ of $Sp(4)$ in an
$SU(2) \times U(1)$ basis; there is then a
missing label. There is no real need to
orthonormalize the states.

\let\picnaturalsize=N
\def\picsize{1.0in}
\def\picfilename{mira.eps}
\begin{figure}[p]
\ifx\nopictures Y\else{\ifx\epsfloaded Y\else\input epsf \fi
\global\let\epsfloaded=Y
\centerline{\ifx\picnaturalsize N\epsfxsize \picsize\fi \epsfxsize
5.2 truein \epsfbox{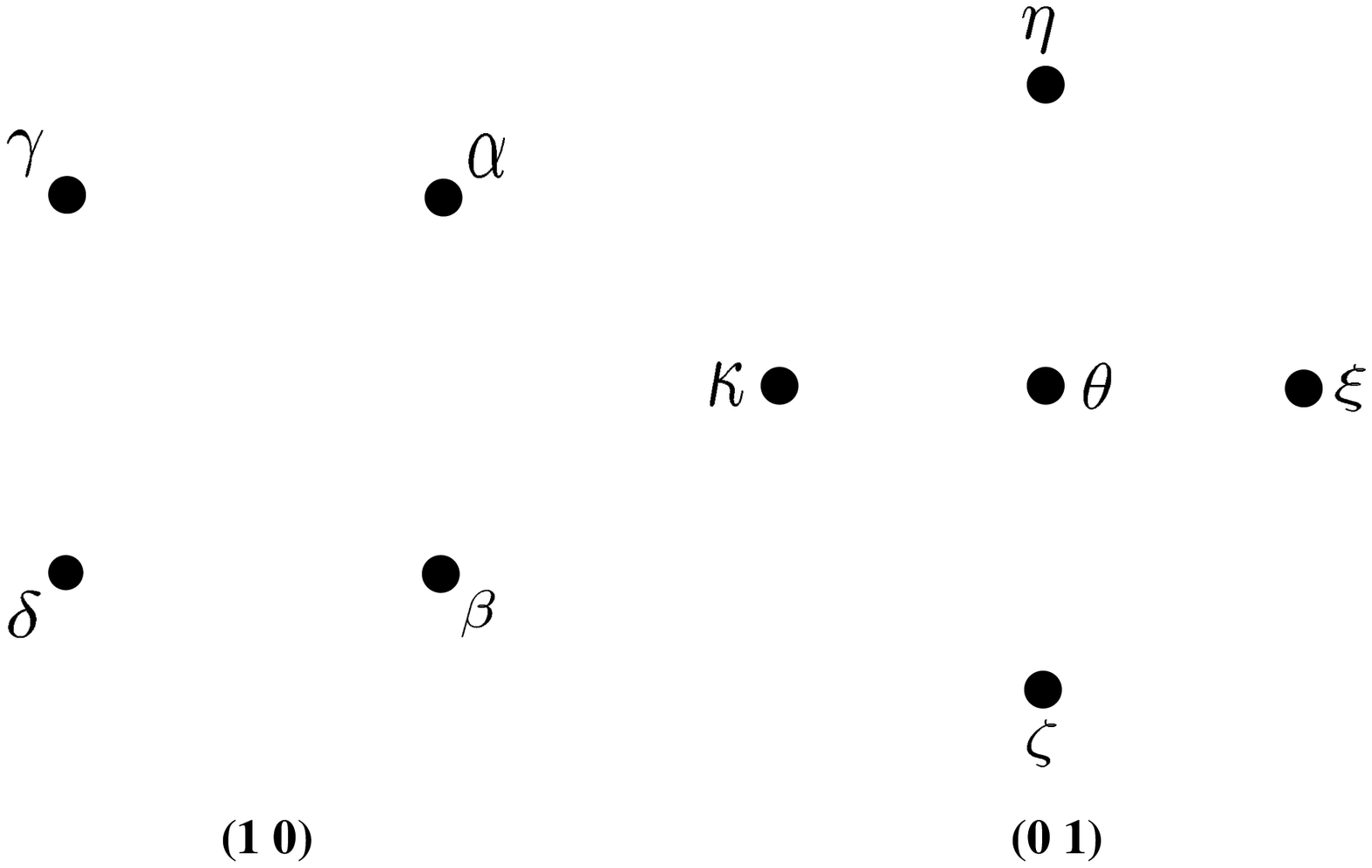}}}\fi
\caption{States of the fundamental representations of $Sp(4)$.}
\end{figure}

\pagebreak

\setcounter{footnote}{0}
\section{ References}

Flowers B H 1952 Studies in $jj-$coupling I
Classification of Nuclear and Atomic States.
{\it Proc. Roy. Soc.} (London) {\bf A212}
248 -- 263
\medskip

Hecht K T 1965 Some Simple $R_5$ Wigner Coefficients
and Their Applications.
{\it Nucl. Phys.} {\bf 63} 177 -- 213
\medskip

Parikh J C 1965 The Role of Isospin Pair
Correlations for Configurations of the
Type $(J)^N$. {\it Nucl. Phys.} {\bf 63}
214 -- 232
\medskip

Ahmed K and Sharp R T 1970 $O(5)$ Bases
for Nuclear Seniority Model.
{\it J. Math. Phys.} {\bf 11}
1112 -- 1117
\medskip

Smirnov Y F and Tolstoy V N 1973
A New Projected Basis in the Theory
of Five--Dimensional Quasi--Spin.
{\it Rep. Math. Phys.} {\bf 4} 97 -- 111
\medskip

de Guise H and Sharp R T 1991 Polynomial
States for $SU(3)$ and $SO(5)$ in a
Demazure--Tits basis. {\it J. Phys. A: Math. Gen.}
{\bf 24} 557 -- 568
\medskip

Burdik {\v C}, Cummins C J, Gaskell R W and
Sharp R T 1992 Complete Branching
Rules Generating Function for
$SO(7) \supset SU(2)^3$ and Polynomial
Basis States. {\it J. Phys. A: Math. Gen.}
4835 -- 4846
\medskip

Farell L, Lam C S and Sharp R T
1994 $G_2$ Generator Matrix Elements for Degenerate
Representations in an $SU(3)$ Basis: I.
{\it J. Phys. A: Math. Gen.} {\bf 27} 2761 -- 2771
\medskip

Hambli N and Sharp R T 1995 Generator Matrix Elements
for $G_2 \supset SU(3)$: II. Generic Representation.
{\it J. Phys. A: Math. Gen.} {\bf 28} 2581 -- 2588
\medskip

Gaskell R, Peccia A and Sharp R T
1978 Generating Functions for Polynomial
Irreducible Tensors. {\it J. Math. Phys.}
{\bf 19} 727 -- 733
\medskip

Patera J and Sharp R T 1979 Generating
Functions for Characters of Group Representations
and Their Applications Lecture Notes in Physics
{\bf 94} 175 -- 183
\medskip

\end{document}